\documentclass[aps, prd, amsfonts, amssymb, amsmath,
12pt, onecolumn, notitlepage, nofootinbib, showpacs]{revtex4-1}
\usepackage{graphicx}
\newcommand{\DP}{\Delta\Pi}
\newcommand{\ind}[2]{^{\text{\scriptsize $#1$}}_{\text{\scriptsize #2}}}
\newcommand{\inds}[2]{^{\text{\scriptsize $#1$}}_{\text{\tiny #2}}}
\newcommand{\Nc}{N_{\text{\scriptsize c}}}
\newcommand{\nf}{n_{\text{\scriptsize f}}}
\newcommand{\nfs}{n_{\text{\tiny f}}}
\newcommand\AL[1]{\Delta\alpha\ind{#1}{lep}}
\newcommand\AH[1]{\Delta\alpha\ind{#1}{had}}
\newcommand\AMH[1]{\Delta\alpha\ind{#1}{had}}

\begin{document}

\title{Hadronic vacuum polarization function \\ within dispersive approach to QCD}

\author{A.V.~Nesterenko}
\email{nesterav@theor.jinr.ru}
\affiliation{Bogoliubov Laboratory of Theoretical Physics,
Joint Institute for Nuclear Research,
Dubna, 141980, Russian Federation}

\begin{abstract}
The dispersive approach to quantum chromodynamics is applied to the study
of the hadronic vacuum polarization function and associated quantities.
This approach merges the intrinsically nonperturbative constraints, which
originate in the kinematic restrictions on the respective physical
processes, with corresponding perturbative input. The obtained hadronic
vacuum polarization function agrees with pertinent lattice simulation
data. The evaluated hadronic contributions to the muon anomalous magnetic
moment and to the shift of the electromagnetic fine structure constant
conform with recent assessments of these quantities.
\end{abstract}



\maketitle

\section{Introduction}
\label{Sect:Intro}

The theoretical description of a number of the strong interaction
processes is inherently based on the hadronic vacuum polarization
function~$\Pi(q^2)$. In particular, this function plays a crucial role in
the studies of inclusive $\tau$~lepton hadronic decay and of
electron--positron annihilation into hadrons, that provides decisive
self--consistency tests of quantum chromodynamics~(QCD). At the same time,
the function~$\Pi(q^2)$ enters in the analysis of the hadronic
contributions to such quantities of precise particle physics as the muon
anomalous magnetic moment and the running of the electromagnetic fine
structure constant, that, in turn, puts strong limits on the effects due
to a possible new fundamental physics beyond the standard model~(SM).
Additionally, the theoretical exploration of the aforementioned processes
constitutes a natural framework for a thorough investigation of both
perturbative and intrinsically nonperturbative aspects of hadron dynamics.

The strong interactions possess the feature of the asymptotic freedom,
that makes it possible to apply perturbation theory to the study of
ultraviolet behaviour of the function~$\Pi(q^2)$. However, there is still
no rigorous method of theoretical description of hadron dynamics at low
energies, which would have provided one with robust unabridged results.
This fact eventually forces one to engage a variety of nonperturbative
approaches in order to examine the strong interactions in the infrared
domain. For example, an insight into the low--energy behaviour of the
hadronic vacuum polarization function can be gained from such methods as,
e.g., lattice simulations~\cite{Lat1, Lat2, Lat3, Lat4}, operator product
expansion~\cite{OPE1, OPE2, OPE3, OPE4, OPE5, OPE6}, instanton liquid
model~\cite{NLCQM, ILM}, and others.

Theoretical particle physics widely employs various methods based on the
dispersion relations\footnote{Among the recent applications of such
methods are, for example, the extension of applicability range of chiral
perturbation theory~\cite{Portoles, Passemar}, the precise determination
of parameters of resonances~\cite{Kaminski}, the assessment of the
hadronic light--by--light scattering~\cite{DispHlbl}, and many
others~\cite{APT, APT1, APT2, APT3, APT4, APT5, APT6, APT7a, APT7b, APT8,
APT9, APT10, APT11}.}. In~particular, the latter provide a source of the
nonperturbative information about the low--energy hadron dynamics.
Specifically, the dispersion relations, which render the kinematic
restrictions on the relevant physical processes into the mathematical
form, impose stringent constraints on the pertinent quantities [such
as~$\Pi(q^2)$ and related functions], that should certainly be accounted
for when one comes out of the applicability range of perturbation theory.
These nonperturbative constraints have been merged with corresponding
perturbative input in the framework of dispersive approach to
QCD~\cite{DQCD1a, PRD88}, which provides unified integral representations
for the functions on hand, see Sec.~\ref{Sect:DQCD}.

The primary objective of this paper is to calculate the hadronic vacuum
polarization function within dispersive approach and to compare it with
relevant lattice simulation data, as well as to evaluate the corresponding
hadronic contributions to the muon anomalous magnetic moment and to the
shift of the electromagnetic fine structure constant.

The layout of the paper is as follows. In Sec.~\ref{Sect:DQCD} the
dispersive approach to QCD~\cite{DQCD1a, PRD88} is overviewed.
Section~\ref{Sect:Latt} presents the comparison of the hadronic vacuum
polarization function calculated in the framework of dispersive approach
with pertinent lattice simulation data and elucidates the qualitative
distinctions between the approach on hand, its massless limit, and
perturbative approach. Section~\ref{Sect:EW} contains the evaluation of
hadronic contributions to the aforementioned electroweak observables.
In~the Conclusions (Sect.~\ref{Sect:Concl}) the basic results are
summarized and further studies within this approach are outlined.
Auxiliary material is given in the~Appendix.

\section{Dispersive approach to Quantum Chromodynamics}
\label{Sect:DQCD}

The hadronic vacuum polarization function~$\Pi(q^2)$ is defined as the
scalar part of the hadronic vacuum polarization tensor
\begin{equation}
\label{P_Def}
\Pi_{\mu\nu}(q^2) = i\!\int\!d^4x\,e^{i q x} \langle 0 |\,
T\!\left\{J_{\mu}(x)\, J_{\nu}(0)\right\} | 0 \rangle =
\frac{i}{12\pi^2} (q_{\mu}q_{\nu} - g_{\mu\nu}q^2) \Pi(q^2).
\end{equation}
The kinematics of the process on hand determines the cut structure
of~$\Pi(q^2)$ in the complex $q^2$--plane. Specifically, the
function~$\Pi(q^2)$~(\ref{P_Def}) has the only cut along the positive
semiaxis of real~$q^2$ starting at the hadronic production
threshold~$4m_{\pi}^2=m^2$ (discussion of this issue can be found in,
e.g., Ref.~\cite{Feynman}, as well as in Refs.~\cite{DQCD1a, PRD88,
DQCDPrelim1}). Proceeding from this fact and bearing in mind the
asymptotic ultraviolet behaviour of the hadronic vacuum polarization
function one can write down the corresponding dispersion relation by
making use of the once--subtracted Cauchy integral formula [see
Eq.~(\ref{P_Disp}) below]. For practical purposes it proves to be
convenient to define the Adler function~$D(Q^2)$~\cite{Adler} [see
Eq.~(\ref{Adler_Def}) below] and the related function~$R(s)$, which is
identified with the so--called $R$--ratio of electron--positron
annihilation into hadrons [see Eq.~(\ref{R_Def}) below]. Eventually, the
complete set of well--known relations, which express the
functions~$\Pi(q^2)$, $R(s)$, and~$D(Q^2)$ in terms of each other,
acquires the following form (see papers~\cite{Adler, PDisp, RKP82} as well
as~\cite{PRD88} and references therein):
\begin{eqnarray}
\label{P_Disp}
\DP(q^2\!,\, q_0^2) &=& (q^2 - q_0^2) \int_{m^2}^{\infty}
\frac{R(\sigma)}{(\sigma-q^2)(\sigma-q_0^2)}\, d\sigma \qquad \\[1.5mm]
\label{P_Disp2}
&=& - \int_{-q_0^2}^{-q^2} D(\zeta) \frac{d \zeta}{\zeta},
\end{eqnarray}
\begin{eqnarray}
\label{R_Def}
R(s) &=& \frac{1}{2 \pi i} \lim_{\varepsilon \to 0_{+}}\!
\DP(s+i\varepsilon, s-i\varepsilon) \\[1.5mm]
\label{R_Disp2}
&=&  \frac{1}{2 \pi i} \lim_{\varepsilon \to 0_{+}}
\int_{s + i \varepsilon}^{s - i \varepsilon}
D(-\zeta)\,\frac{d \zeta}{\zeta},
\end{eqnarray}
\begin{eqnarray}
\label{Adler_Def}
D(Q^2) &=& - \frac{d\, \Pi(-Q^2)}{d \ln Q^2} \\[1.5mm]
\label{Adler_Disp}
&=& Q^2 \int_{m^2}^{\infty} \frac{R(\sigma)}{(\sigma+Q^2)^2}\, d\sigma. \qquad
\end{eqnarray}

\begin{figure}[t]
\centerline{\includegraphics[width=90mm]{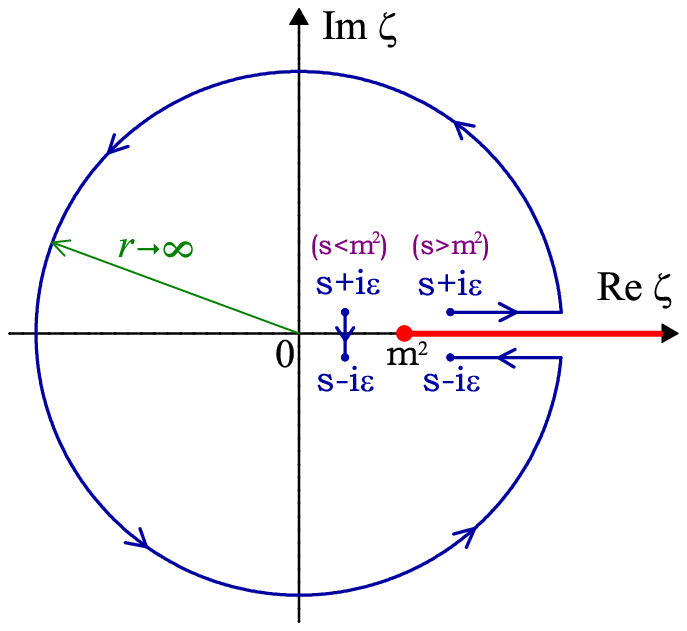}}
\caption{The integration contour in Eq.~(\ref{R_Disp2}). The physical cut
$\zeta \ge m^2$ of the Adler function $D(-\zeta)$~(\ref{Adler_Def}) is
shown along the positive semiaxis of real~$\zeta$.}
\label{Plot:R_Contour}
\end{figure}

In Eqs.~(\ref{P_Disp})--(\ref{Adler_Disp}) $\DP(q^2\!,\, q_0^2) = \Pi(q^2)
- \Pi(q_0^2)$, whereas $Q^2 = -q^2 > 0$ and $s = q^2 > 0$ denote the
spacelike and timelike kinematic variables, respectively. The common
prefactor $\Nc\sum_{f=1}^{\nfs} Q_{f}^{2}$ is omitted throughout the
paper, where $\Nc=3$ is the number of colours, $Q_{f}$~stands for the
electric charge of $f$--th quark (in units of the elementary charge~$e$),
and $\nf$~denotes the number of active flavours. The integration contour
in Eq.~(\ref{R_Disp2}) lies in the region of analyticity of its integrand
(see Fig.~\ref{Plot:R_Contour}). Note that the derivation of
relations~(\ref{P_Disp})--(\ref{Adler_Disp}) requires the knowledge
of the cut structure of hadronic vacuum polarization
function~$\Pi(q^2)$~(\ref{P_Def}) and its asymptotic ultraviolet
behaviour. It is worth mentioning also that Eqs.~(\ref{P_Disp})
and~(\ref{Adler_Disp}) can be used for extracting the
functions~$\DP(q^2\!,\, q_0^2)$ and~$D(Q^2)$ from the experimental data
on~$R(s)$.

\medskip

As noted in the Introduction, the dispersion
relations~(\ref{P_Disp})--(\ref{Adler_Disp}) embody the kinematic
restrictions on the respective physical processes and impose intrinsically
nonperturbative constraints on the functions~$\Pi(q^2)$, $R(s)$,
and~$D(Q^2)$, that should certainly be taken into account when one
oversteps the limits of applicability of perturbation theory. These
nonperturbative constraints\footnote{Including the correct analytic
properties in the kinematic variable, that implies the absence of
unphysical singularities in Eqs.~(\ref{P_DQCD})--(\ref{Adler_DQCD}), see
Sec.~II$\,$A of Ref.~\cite{PRD88} for the details.} have been merged with
corresponding perturbative input in the framework of dispersive approach
to QCD\footnote{Its preliminary formulation was discussed in
Refs.~\cite{DQCDPrelim1, DQCDPrelim2}.}~\cite{DQCD1a, PRD88}, which
provides the following unified integral representations for the functions
on hand:
\begin{eqnarray}
\label{P_DQCD}
\DP(q^2,\, q_0^2) &=& \DP^{(0)}(q^2,\, q_0^2) +
\!\int_{m^2}^{\infty} \rho(\sigma)
\ln\biggl(\frac{\sigma-q^2}{\sigma-q_0^2}
\frac{m^2-q_0^2}{m^2-q^2}\biggr)\frac{d\,\sigma}{\sigma}, \\[1.5mm]
\label{R_DQCD}
R(s) &=& R^{(0)}(s) + \theta(s-m^2) \int_{s}^{\infty}\!
\rho(\sigma) \frac{d\,\sigma}{\sigma}, \\[1.5mm]
\label{Adler_DQCD}
D(Q^2) &=& D^{(0)}(Q^2) + \frac{Q^2}{Q^2+m^2}
\int_{m^2}^{\infty} \rho(\sigma)
\frac{\sigma-m^2}{\sigma+Q^2} \frac{d\,\sigma}{\sigma}.
\end{eqnarray}
These equations have been obtained by employing the
relations~(\ref{P_Disp})--(\ref{Adler_Disp}) and the asymptotic
ultraviolet behaviour of the hadronic vacuum polarization function.
In~Eqs.~(\ref{P_DQCD})--(\ref{Adler_DQCD}) $\rho(\sigma)$ is the spectral
density
\begin{eqnarray}
\label{RhoGen}
\rho(\sigma) &=& \frac{1}{2 \pi i} \frac{d}{d\,\ln\sigma}
\lim_{\varepsilon \to 0_{+}}
\Bigl[p(\sigma-i\varepsilon)-p(\sigma+i\varepsilon) \Bigr]
\nonumber \\[1.5mm]
&=& - \frac{d}{d\,\ln\sigma}\, r(\sigma) \\[1.5mm]
&=& \frac{1}{2 \pi i} \lim_{\varepsilon \to 0_{+}}
\Bigl[d(-\sigma-i\varepsilon)-d(-\sigma+i\varepsilon) \Bigr]\!,
\nonumber
\end{eqnarray}
$p(q^2)$, $r(s)$, and~$d(Q^2)$ denote the strong corrections to the
functions~$\Pi(q^2)$, $R(s)$, and~$D(Q^2)$, respectively, $\theta(x)$
stands for the unit step--function [$\theta(x)=1$ if $x \ge 0$ and
$\theta(x)=0$ otherwise], and the leading--order terms read~\cite{Feynman,
QEDAB}:
\begin{eqnarray}
\label{P0L}
\DP^{(0)}(q^2,\, q_0^2) &=& 2\,\frac{\varphi - \tan\varphi}{\tan^3\varphi}
- 2\,\frac{\varphi_{0} - \tan\varphi_{0}}{\tan^3\varphi_{0}}, \\[1.5mm]
\label{R0L}
R^{(0)}(s) &=& \theta(s - m^2)\biggl(1-\frac{m^2}{s}\biggr)^{\!\!3/2}, \\[1.5mm]
\label{D0L}
D^{(0)}(Q^2) &=& 1 + \frac{3}{\xi}\Bigl[1 - \sqrt{1 + \xi^{-1}}\,
\sinh^{-1}\!\bigl(\xi^{1/2}\bigr)\Bigr]\!,
\end{eqnarray}
where $\sin^2\!\varphi = q^2/m^2$, $\sin^2\!\varphi_{0} = q^{2}_{0}/m^2$,
and $\xi=Q^2/m^2$, see papers~\cite{DQCD1a, PRD88} and references therein
for the details.

It~is worthwhile to outline that the Adler function obtained in the
framework of the dispersive approach\footnote{The studies of the Adler
function within other approaches can be found in, e.g., Refs.~\cite{MSS,
Cvetic, Maxwell, Kataev, Fischer, PeRa, BJ}.}~(\ref{Adler_DQCD}) agrees
with its experimental prediction in the entire energy range, see
Refs.~\cite{DQCD1a, DQCD1b, DQCD2}. At the same time, the
representations~(\ref{P_DQCD})--(\ref{Adler_DQCD}) conform with the
results of Bethe--Salpeter calculations~\cite{PRL99PRD77} as well as of
lattice simulations~\cite{RCTaylor}. Additionally, the dispersive approach
has proved to be capable of describing OPAL (update~2012,
Ref.~\cite{OPAL9912}) and ALEPH (update~2014, Ref.~\cite{ALEPH0514})
experimental data on inclusive $\tau$~lepton hadronic decay in vector and
axial--vector channels in a self--consistent way~\cite{PRD88, QCD14} (see
also Refs.~\cite{DQCD3, C12}).

In the framework of the approach on hand the corresponding perturbative
input is accounted for in the same way as in other similar approaches,
namely, by means of the spectral density~(\ref{RhoGen}). Specifically, the
latter is approximated by its perturbative part, which can be calculated
by making use of the perturbative expression for either of the strong
corrections $p(q^2)$, $r(s)$, and~$d(Q^2)$, see, e.g., papers~\cite{CPC,
BCmath} and references therein:
\begin{eqnarray}
\label{RhoPert}
\rho\ind{}{pert}(\sigma) &=& \frac{1}{2 \pi i} \frac{d}{d\,\ln\sigma}
\lim_{\varepsilon \to 0_{+}}
\Bigl[p\ind{}{pert}(\sigma-i\varepsilon) -
p\ind{}{pert}(\sigma+i\varepsilon) \Bigr]
\nonumber \\[1.5mm]
&=&  - \frac{d}{d\,\ln\sigma}\, r\ind{}{pert}(\sigma) \\[1.5mm]
&=& \frac{1}{2 \pi i} \lim_{\varepsilon \to 0_{+}}
\Bigl[d\ind{}{pert}(-\sigma-i\varepsilon) -
d\ind{}{pert}(-\sigma+i\varepsilon) \Bigr]\!.
\nonumber
\end{eqnarray}
It is worth noting here that in the massless limit $(m^2 = 0)$ for the
case of perturbative spectral function~(\ref{RhoPert}) Eqs.~(\ref{R_DQCD})
and~(\ref{Adler_DQCD}) become identical to those of the analytic
perturbation theory~(APT)~\cite{APT} (see also Refs.~\cite{APT1, APT2,
APT3, APT4, APT5, APT6, APT7a, APT7b, APT8, APT9, APT10, APT11}). However,
as discussed in Refs.~\cite{PRD88, DQCD1a, C12, DQCD2}, the massless limit
loses some of the substantial nonperturbative constraints, which relevant
dispersion relations impose on the functions on hand, that appears to be
essential for the studies of hadron dynamics at low energies.

\section{Comparison of~$\Pi(q^2)$ with lattice simulation data}
\label{Sect:Latt}

As mentioned above, the lattice QCD simulations constitute an efficient
method of investigation of the nonperturbative aspects of strong
interactions. Over past time this method has been applied to an extensive
study of a broad range of topics (for a recent overview see, e.g.,
Ref.~\cite{LattRev}), including the low--energy behaviour of the hadronic
vacuum polarization function~$\Pi(q^2)$. It is of a particular interest to
juxtapose the function~$\Pi(q^2)$ obtained within dispersive
approach~(\ref{P_DQCD}) with relevant lattice simulation data.

To calculate the hadronic vacuum polarization function, it is convenient
to proceed with the subtracted at zero form of Eq.~(\ref{P_DQCD}), namely
\begin{equation}
\label{P_DQCD2}
\bar{\Pi}(Q^2) = \DP(0,-Q^2) = \DP^{(0)}(0, -Q^2) +
\int_{m^2}^{\infty} \rho(\sigma)
\ln\biggl(\frac{1+Q^2/m^2}{1+Q^2/\sigma}\biggr)
\frac{d\,\sigma}{\sigma}.
\end{equation}
In what follows we shall employ the perturbative expression for the
spectral function~(\ref{RhoPert}). At the one--loop level it assumes a
simple form [namely, $\rho\ind{(1)}{pert}(\sigma) = (4/\beta_{0})
[\ln^{2}(\sigma/\Lambda^2)+\pi^2]^{-1}$, where $\beta_{0} = 11 - 2\nf/3$
and~$\Lambda$ denotes the QCD scale parameter], whereas at the higher loop
levels~$\rho\ind{}{pert}(\sigma)$~(\ref{RhoPert}) is rather cumbrous. The
explicit expressions for the spectral function~(\ref{RhoPert}) up to the
four--loop level\footnote{Recently completed calculation of the respective
four--loop perturbative coefficient is given in Ref.~\cite{AdlerPert4L}.}
can be found in Ref.~\cite{CPC}.

\begin{figure}[t]
\centerline{\includegraphics[width=90mm]{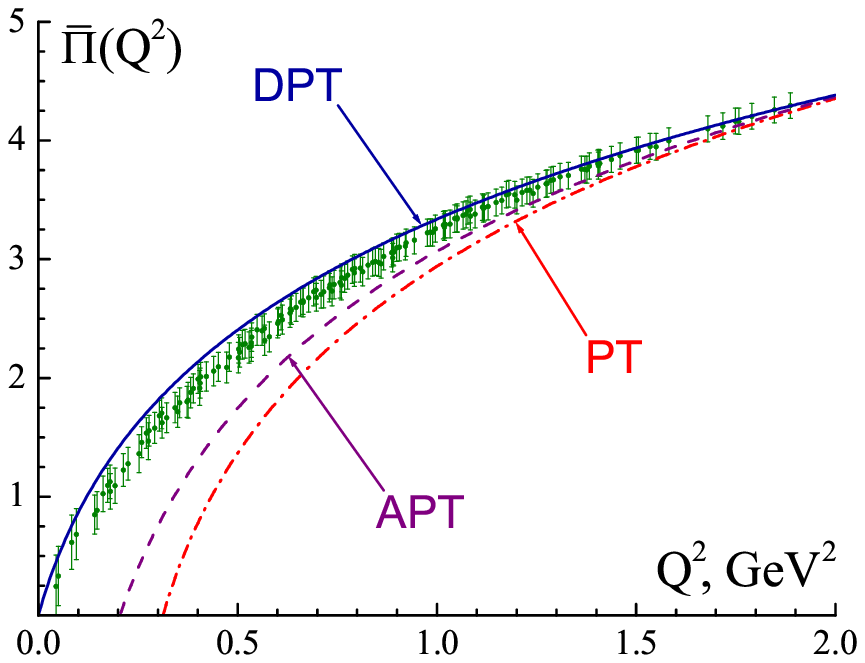}}
\caption{Comparison of the four--loop hadronic vacuum polarization
function calculated within dispersive approach~(\ref{P_DQCD2}) (solid
curve) with lattice simulation data~\cite{Lat5} (circles). The massless
prediction of~$\Pi(q^2)$~(\ref{PAPT1L}) is denoted by dashed curve,
whereas its perturbative approximation~(\ref{PPert1L}) is shown by
dot--dashed curve. Values of parameters: $\Lambda=419\,$MeV, $\nf=2$.}
\label{Plot:PDQCD}
\end{figure}

As one can infer from Fig.~\ref{Plot:PDQCD}, the hadronic vacuum
polarization function~(\ref{P_DQCD2}) (solid curve), which was obtained
within dispersively improved perturbation theory~(DPT) delineated in the
previous Section, is in a good agreement with lattice simulation
data~\cite{Lat5}~(circles) (the rescaling procedure described in
Refs.~\cite{RhoRescale1, RhoRescale2} was applied). The presented result
corresponds to the four--loop level and $\nf=2$~active flavours. To
elucidate the qualitative distinctions between the approaches mentioned in
the previous Section, Fig.~\ref{Plot:PDQCD} also displays the one--loop
Eq.~(\ref{P_DQCD}) in the massless limit, which, in the considered case,
corresponds to APT (dashed curve)
\begin{equation}
\label{PAPT1L}
\DP\inds{(1)}{APT}(-Q^{2},-Q_{0}^{2}) = \DP\ind{(0)}{pert}(-Q^2,\, -Q_0^2)
- \frac{4}{\beta_{0}}
\ln\!\Biggl[\frac{a\ind{(1)}{an}(Q_{0}^{2})}{a\ind{(1)}{an}(Q^{2})}\Biggr]\!,
\end{equation}
and the one--loop perturbative approximation of~$\Pi(q^2)$ (dot--dashed
curve)
\begin{equation}
\label{PPert1L}
\DP\ind{(1)}{pert}(-Q^{2},-Q_{0}^{2}) = \DP\ind{(0)}{pert}(-Q^2,\, -Q_0^2)
- \frac{4}{\beta_{0}}
\ln\!\Biggl[\frac{a\ind{(1)}{pert}(Q_{0}^{2})}{a\ind{(1)}{pert}(Q^{2})}\Biggr]\!.
\end{equation}
In these equations the leading--order terms read
\begin{equation}
\label{P_Pert0}
\DP\ind{(0)}{pert}(-Q^2,\, -Q_0^2) =
-\ln\biggl(\frac{Q^2}{Q_0^2}\biggr)\!,
\end{equation}
the notation $a(Q^2)=\alpha(Q^2)\beta_{0}/(4\pi)$ is used,
\begin{equation}
\alpha\ind{(1)}{pert}(Q^2) =
\frac{4\pi}{\beta_{0}}\,\frac{1}{\ln z}, \qquad z=\frac{Q^2}{\Lambda^2}
\end{equation}
denotes the one--loop perturbative running coupling, and
\begin{equation}
\label{NAIC}
\alpha\ind{(1)}{an}(Q^2) =
\frac{4\pi}{\beta_{0}}\,\frac{z-1}{z\,\ln z}
\end{equation}
stands for the one--loop infrared enhanced analytic running coupling. It
is interesting to note here that the expression~(\ref{NAIC}) was first
obtained in Refs.~\cite{PRD6264, Review} and has been independently
rediscovered (proceeding from entirely different reasoning) later on in
Ref.~\cite{Schrempp}, see also Ref.~\cite{RCProsperi}.

As mentioned above, the dispersion
relations~(\ref{P_Disp})--(\ref{Adler_Disp}) impose intrinsically
nonperturbative constraints on the functions on hand, whereas the integral
representations~(\ref{P_DQCD})--(\ref{Adler_DQCD}) merge these constraints
with corresponding perturbative input. For~example, as discussed in
Refs.~\cite{DQCD1a, PRD88}, the relation~(\ref{Adler_Disp}) implies that
the Adler function~$D(Q^2)$ possesses the only cut~$Q^2 \le -m^2$ along
the negative semiaxis of real~$Q^2$ and that~$D(Q^2)$ vanishes in the
infrared limit~$Q^2 \to 0$ (this condition holds for~$m \neq 0$ only and
appears to be lost in the massless limit). In~turn, the first of these
constraints indicates that the Adler function~(\ref{Adler_DQCD}) contains
no unphysical singularities, whereas the second one substantially
stabilizes its infrared behaviour, see Refs.~\cite{DQCD1a, PRD88}.
Similarly, relation~(\ref{P_Disp}) signifies that the hadronic vacuum
polarization function~$\Pi(q^2)$ possesses the only cut~$q^2 \ge m^2$
along the positive semiaxis of real~$q^2$ and that the subtraction
point~$q_{0}^{2}$ can be located anywhere in the complex~$q^2$--plane
except for this~cut. In~turn, the first of these constraints means that
the hadronic vacuum polarization function~(\ref{P_DQCD}) is free of the
unphysical singularities, whereas the second one enables one to
subtract~$\Pi(q^2)$ at~$q_{0}^{2} = 0$ (for $m \neq 0$ only), that binds
the low--energy behaviour of~$\bar{\Pi}(Q^2)$~(\ref{P_DQCD2}). This issue
is illustrated by Fig.~\ref{Plot:PDQCD}. Specifically, the perturbative
approximation of the hadronic vacuum polarization function~(\ref{PPert1L})
(dot--dashed curve) contains infrared unphysical singularities, that makes
it inapplicable at low energies. At~the same time, although both
expressions~(\ref{P_DQCD2}) and~(\ref{PAPT1L}) are free of the unphysical
singularities, their infrared behaviour is quite different. Namely, the
hadronic vacuum polarization function~(\ref{P_DQCD2}) (solid curve)
vanishes in the infrared limit, whereas the APT prediction~(\ref{PAPT1L})
diverges at~$Q^2 \to 0$. The latter originates in the mathematical fact
that in the massless limit the function~$\Pi(q^2)$ is undefined at the
beginning of its branch~cut. This makes the massless APT prediction
of~$\Pi(q^2)$ also incompatible with lattice simulation data at low
energies. It~is worthwhile to note also that the aforementioned features
are universal and determine the qualitative behaviour of the hadronic
vacuum polarization function within each of the approaches discussed
above.

\section{Hadronic contributions to electroweak observables}
\label{Sect:EW}

\subsection{Muon anomalous magnetic moment}
\label{Sect:Muon}

The theoretical description of the muon anomalous magnetic moment $a_{\mu}
= (g_{\mu}-2)/2$ is a long--standing challenging issue of the elementary
particle physics, which engages the entire pattern of interactions
within~SM. Both experimental measurements~\cite{MuonExp1, MuonExp2} and
theoretical evaluations~\cite{MuonRev1, MuonRev2} of~$a_{\mu}$ have
achieved an impressive accuracy, and the remaining discrepancy of the
order of few standard deviations between them may be an evidence for the
existence of a new physics beyond~SM. The uncertainty of theoretical
estimation of~$a_{\mu}$ is mainly dominated by the leading--order hadronic
contribution~$a_{\mu}^{\mbox{\tiny HLO}}$, which involves the integration
of hadronic vacuum polarization function~$\Pi(q^2)$ over the range
inaccessible within perturbation theory\footnote{To obviate this
difficulty one can express $a_{\mu}^{\mbox{\tiny HLO}}$~(\ref{AmuHVP}) in
terms of~$R(s)$ by making use of Eq.~(\ref{P_Disp}) and replace the
low--energy behaviour of~$R(s)$ with relevant experimental data, see
reviews~\cite{MuonRev1, MuonRev2} and references therein.} (see, e.g.,
Ref.~\cite{Raf72}):
\begin{equation}
\label{AmuHVP}
a_{\mu}^{\mbox{\tiny HLO}} = \frac{1}{3} \biggl(\frac{\alpha}{\pi}\biggr)^{\!\!2}
\!\int_{0}^{\infty}\!\! f\!\biggl(\frac{\zeta}{4m_{\mu}^{2}}\biggr)
\bar{\Pi}(\zeta) \frac{d\zeta}{4m_{\mu}^{2}}
= \frac{1}{3} \biggl(\frac{\alpha}{\pi}\biggr)^{\!\!2}
\!\int_{0}^{1}\!(1-x)
\bar{\Pi}\!\biggl(\!m_{\mu}^{2}\,\frac{x^2}{1-x}\!\biggr) dx.
\end{equation}
In this equation
\begin{equation}
f(x) = \frac{1}{x^3}\,\frac{y^{5}(x)}{1-y(x)}
\end{equation}
and $y(x) = x\bigl(\sqrt{1+x^{-1}} - 1\bigr)$ is a monotonously
nondecreasing function of its argument, $0 \le y(x) < 1/2$.

\begin{figure}[t]
\centerline{\includegraphics[width=90mm,clip]{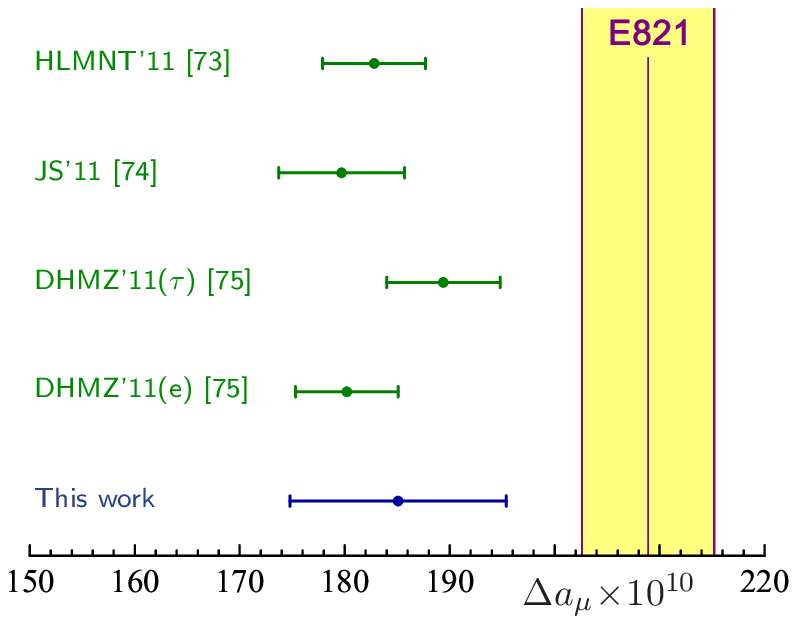}}
\caption{Comparison of the subtracted muon anomalous magnetic
moment~(\ref{Amu_DQCD}) with its recent assessments~\cite{HLMNT11, JS11,
DHMZ11} (circles). The averaged experimental value~(\ref{Amu_Exp}) is
shown by vertical shaded band, $\Delta a_{\mu} = a_{\mu} - a_{0}$, and
$a_{0} = 11659 \times 10^{-7}$.}
\label{Plot:Amu}
\end{figure}

As mentioned above, in the framework of dispersive approach the hadronic
vacuum polarization function~$\Pi(q^2)$~(\ref{P_DQCD}) is free of the
unphysical singularities, that enables one to perform the integration in
Eq.~(\ref{AmuHVP}) in a straightforward way [i.e.,~without involving
experimental data on~$R(s)$]. Thus, to evaluate~$a_{\mu}^{\mbox{\tiny
HLO}}$ within approach on hand, we shall employ Eqs.~(\ref{AmuHVP})
and~(\ref{P_DQCD2}) with the four--loop spectral function~(\ref{RhoPert}),
that eventually results~in
\begin{equation}
\label{AmuHLO_DQCD}
a_{\mu}^{\mbox{\tiny HLO}} = (696.1 \pm 9.5) \times 10^{-10}.
\end{equation}
In this equation the quoted error accounts for the uncertainties of the
parameters entering Eq.~(\ref{AmuHVP}), their values being taken from
Refs.~\cite{PDG2012, CODATA2012}. The obtained estimation of the
leading--order hadronic contribution to~$a_\mu$~(\ref{AmuHLO_DQCD})
appears to be in a good agreement with its recent assessments, namely,
$a_{\mu}^{\mbox{\tiny HLO}} = (694.9 \pm 4.3) \times 10^{-10}$
(Ref.~\cite{HLMNT11}), $a_{\mu}^{\mbox{\tiny HLO}} = (691.0 \pm 4.7)
\times 10^{-10}$ (Ref.~\cite{JS11}), $a_{\mu}^{\mbox{\tiny HLO}} = (701.5
\pm 4.7) \times 10^{-10}$ ($\tau$--based) and $a_{\mu}^{\mbox{\tiny HLO}}
= (692.3 \pm 4.2) \times 10^{-10}$ ($e^{+}e^{-}$--based)
(Ref.~\cite{DHMZ11}).

To evaluate the complete SM prediction of the muon anomalous magnetic
moment~$a_{\mu}$ one has also to account for the QED contribution
$a_{\mu}^{\mbox{\tiny QED}} = (11658471.8951 \pm 0.0080) \times
10^{-10}$~\cite{AmuQED}, the electroweak contribution
$a_{\mu}^{\mbox{\tiny EW}} = (15.36 \pm 0.10) \times
10^{-10}$~\cite{AmuEW}, as well as the higher--order $a_{\mu}^{\mbox{\tiny
HHO}} = (-9.84 \pm 0.07) \times 10^{-10}$~\cite{HLMNT11} and
light--by--light $a_{\mu}^{\mbox{\tiny Hlbl}} = (11.6 \pm 4.0) \times
10^{-10}$~\cite{AmuHlbl} hadronic contributions, that, together with
$a_{\mu}^{\mbox{\tiny HLO}}$~(\ref{AmuHLO_DQCD}), leads to
\begin{equation}
\label{Amu_DQCD}
a_{\mu} = (11659185.1 \pm 10.3) \times 10^{-10}.
\end{equation}
The difference between this value and the Brookhaven~E821 experimental
measurement\footnote{The averaged experimental value~(\ref{Amu_Exp})
accounts for the recently updated ratio of the muon--to--proton magnetic
moment, see Ref.~\cite{MuonExp3}.}~\cite{MuonExp2}
\begin{equation}
\label{Amu_Exp}
a_{\mu}^{\mbox{\scriptsize exp}} = (11659208.9 \pm 6.3) \times 10^{-10}
\end{equation}
is $(23.8 \pm 12.1)\times 10^{-10}$, that corresponds to the discrepancy
of two~standard deviations. As~one can infer from Fig.~\ref{Plot:Amu},
the estimation of the muon anomalous magnetic
moment~$a_{\mu}$~(\ref{Amu_DQCD}) fairly agrees with its recent
evaluations~\cite{HLMNT11, JS11, DHMZ11}.

\subsection{Electromagnetic fine structure constant}

The electromagnetic running coupling~$\alpha\ind{}{em}(q^2)$ plays a
central role in a variety of issues of precision particle physics. The
vacuum polarization effects screen the electric charge and make the
electromagnetic coupling~$\alpha\ind{}{em}$ dependent on the energy
scale~$q^2$:
\begin{equation}
\label{RC_QED}
\alpha\ind{}{em}(q^2) = \frac{\alpha}{1 - \AL{}(q^2) - \AH{}(q^2)},
\end{equation}
with $\alpha = e^2/(4\pi) \simeq 1/137.036$ being the fine structure
constant. In~Eq.~(\ref{RC_QED}) the leptonic contribution~$\AL{}(q^2)$ can
reliably be calculated by making use of perturbation theory~\cite{AEMLep}.
However, similarly to the aforementioned case of the muon anomalous
magnetic moment, the hadronic contribution to Eq.~(\ref{RC_QED})
\begin{equation}
\label{AMH}
\AMH{}(q^2) = - \frac{\alpha}{3\pi}\,q^2
\,\,\mathcal{P}\!\!\int_{m^2}^{\infty}\!\!
\frac{R(s)}{s-q^2}\,\frac{d\,s}{s}
\end{equation}
($\mathcal{P}$~stands for the ``Cauchy principal value'') involves the
integration over the low--energy range and constitutes the prevalent
source of the uncertainty of~$\alpha\ind{}{em}(q^2)$, see discussion of
this issue in, e.g., papers~\cite{HLMNT11, Passera} and references
therein.

\begin{figure}[t]
\centerline{\includegraphics[width=90mm,clip]{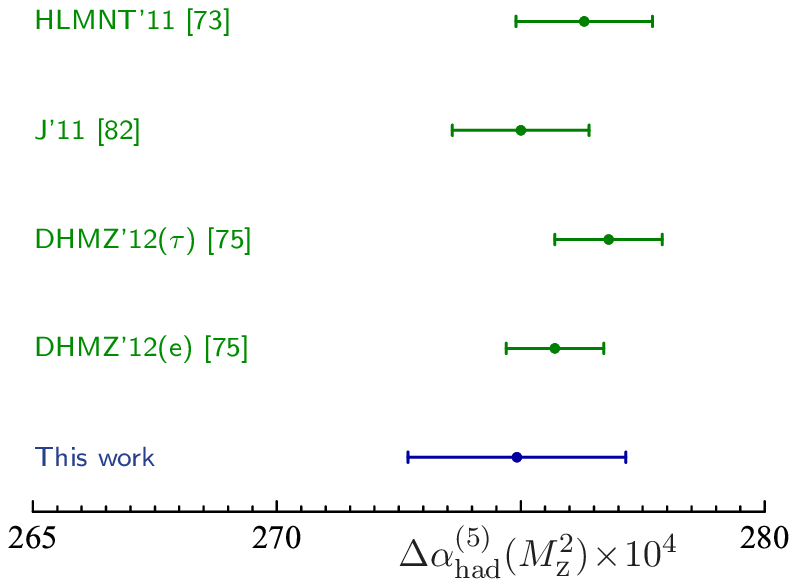}}
\caption{Comparison of the hadronic contribution to the shift of the
electromagnetic fine structure constant at the scale of $Z$~boson
mass~(\ref{AMHDQCD}) with its recent evaluations~\cite{HLMNT11, DHMZ11,
J11}.}
\label{Plot:AHMZ}
\end{figure}

To evaluate the five--flavour\footnote{The respective contribution of the
top quark is, as usual, added separately, see Ref.~\cite{AEMtop}.}
hadronic contribution to the shift of the electromagnetic fine structure
constant at the scale of $Z$~boson mass in the framework of dispersive
approach we shall follow the same lines as in Sec.~\ref{Sect:Muon}, that
eventually yields
\begin{equation}
\label{AMHDQCD}
\AMH{(5)}(M\inds{2}{Z}) = (274.9 \pm 2.2) \times 10^{-4}.
\end{equation}
This equation corresponds to the four--loop level and the quoted error
accounts for the uncertainties of the parameters entering Eq.~(\ref{AMH}),
their values being taken from Refs.~\cite{PDG2012, CODATA2012}. The
obtained estimation of~$\AMH{(5)}(M\inds{2}{Z})$~(\ref{AMHDQCD}) appears
to be in a good agreement with its recent evaluations, specifically,
$\AMH{(5)}(M\inds{2}{Z}) = (276.3 \pm 1.4) \times 10^{-4}$
(Ref.~\cite{HLMNT11}), $\AMH{(5)}(M\inds{2}{Z}) = (275.0 \pm 1.4) \times
10^{-4}$ (Ref.~\cite{J11}), $\AMH{(5)}(M\inds{2}{Z}) = (276.8 \pm 1.1)
\times 10^{-4}$ (\mbox{$\tau$--based}) and $\AMH{(5)}(M\inds{2}{Z}) =
(275.7 \pm 1.0) \times 10^{-4}$ ($e^{+}e^{-}$--based)
(Ref.~\cite{DHMZ11}), see~Fig.~\ref{Plot:AHMZ}. At the same time,
Eq.~(\ref{AMHDQCD}) together with leptonic $\AL{}(M\inds{2}{Z}) = (314.979
\pm 0.002)\times 10^{-4}$~\cite{AEMLep} and top~quark
$\AH{\mbox{top}}(M\inds{2}{Z}) = (-0.70 \pm 0.05) \times
10^{-4}$~\cite{AEMtop} contributions lead to
$\alpha\ind{-1}{em}(M\inds{2}{Z}) = 128.962 \pm 0.030$, that also agrees
with recent assessments of this quantity, namely,
$\alpha\ind{-1}{em}(M\inds{2}{Z}) = 128.962 \pm 0.018$ (Ref.~\cite{J11}),
$\alpha\ind{-1}{em}(M\inds{2}{Z}) = 128.944 \pm 0.019$
(Ref.~\cite{HLMNT11}), and $\alpha\ind{-1}{em}(M\inds{2}{Z}) = 128.952 \pm
0.014$ (Ref.~\cite{DHMZ11}).

It~is worthwhile to mention also that the hadronic parts of the
aforementioned electroweak observables~(\ref{AmuHVP}) and~(\ref{AMH})
receive dominant contributions from different energy ranges. Specifically,
the kernel in Eq.~(\ref{AMH}) signifies that sizable contributions come
from the integration over low and intermediate energies, whereas the
kernel in Eq.~(\ref{AmuHVP}) indicates that a dominant contribution comes
from the integration over the infrared domain. In~particular, the latter
implies that if the experimental data on~$R(s)$ (or~its phenomenological
approximation) are involved into the evaluation of~$a_{\mu}^{\mbox{\tiny
HLO}}$, then the contribution to Eq.~(\ref{AmuHVP}) from the integration
over the range of the lowest lying vector mesons is enhanced. As~one might
also note, the spectral function~(\ref{RhoPert}) contains only
perturbative input. Nonetheless,
$\rho\ind{}{pert}(\sigma)$~(\ref{RhoPert}) appears to be efficient in the
description of the quantities, which can be expressed as convolution
of~$R(s)$ and a~respective kernel over a semi--infinite range.
In~particular, the spectral function~(\ref{RhoPert}) is capable of
describing the Adler function~(\ref{Adler_DQCD}) (see Refs.~\cite{DQCD1a,
PRD88}), the hadronic vacuum polarization function~(\ref{P_DQCD}) (see
Sect.~\ref{Sect:Latt}), and yields the predictions for hadronic
contributions to the aforementioned electroweak observables~(\ref{AmuHVP})
and~(\ref{AMH}), which are in the right ballpark.

\section{Conclusions}
\label{Sect:Concl}

The hadronic vacuum polarization function obtained within dispersive
approach contains no unphysical singularities and agrees with relevant
lattice simulation data. The hadronic contributions to the muon anomalous
magnetic moment and to the shift of the electromagnetic fine structure
constant at the scale of $Z$~boson mass estimated within dispersive
approach conform with recent assessments of these quantities.

In further studies it would undoubtedly be interesting to include into the
presented analysis the nonperturbative contributions arising from the
operator product expansion and to explore possible constraints on the
spectral density appearing in the approach on hand.

\begin{acknowledgments}
The author is grateful to R.~Kaminski, E.~Passemar, M.~Passera,
J.~Portoles, P.~Roig, and H.~Wittig for the stimulating discussions and
useful comments.
\end{acknowledgments}

\appendix*

\section{Correspondence between two sets of relations
for~$\Pi(q^2)$, $R(s)$, and~$D(Q^2)$}
\label{Sect:Corr}

As mentioned in~Ref.~\cite{PRD88}, the integral
representations~(\ref{P_DQCD})--(\ref{Adler_DQCD}) for the
functions~$\Pi(q^2)$, $R(s)$, and~$D(Q^2)$ satisfy all six
relations~(\ref{P_Disp})--(\ref{Adler_Disp}) by construction.
It is straightforward to verify explicitly that the set of
relations~(\ref{P_Disp})--(\ref{Adler_Disp}) holds for the leading--order
terms~\mbox{(\ref{P0L})--(\ref{D0L})} as well as for the most of the
strong corrections~(\ref{P_DQCD})--(\ref{Adler_DQCD}). In particular, to
show that the relations~(\ref{P_Disp2}) and~(\ref{Adler_Def}) are valid
for the pair of the strong
corrections~[(\ref{P_DQCD}),$\,$(\ref{Adler_DQCD})] one has to apply
directly the integration and differentiation, respectively. To demonstrate
that the relations~(\ref{P_Disp}) and~(\ref{Adler_Disp}) hold between
pairs of expressions~[(\ref{P_DQCD}),$\,$(\ref{R_DQCD})]
and~[(\ref{R_DQCD}),$\,$(\ref{Adler_DQCD})] the integration by parts is
required. The validity of relation~(\ref{R_Disp2}) for the
pair~[(\ref{R_DQCD}),$\,$(\ref{Adler_DQCD})] can be shown by employing
\begin{equation}
\lim_{\varepsilon \to 0_{+}} \frac{1}{x \pm i\varepsilon} =
\mp i \pi \delta(x) + \mathcal{P} \frac{1}{x}
\end{equation}
in the respective integrand. The remaining relation~(\ref{R_Def}) between
the pair of the strong corrections~[(\ref{P_DQCD}),$\,$(\ref{R_DQCD})] is
somewhat more laborious to demonstrate than the others, and will be
addressed in this Section.

For the strong corrections~$p(q^2)$ and~$r(s)$ the relation~(\ref{R_Def}) can
be written as
\begin{equation}
\label{Rsc_DQCD}
r(s) = \frac{1}{2 \pi i} \lim_{\varepsilon \to 0_{+}}
\Bigl[\Delta p(s+i\varepsilon, q_{0}^{2}) -
\Delta p(s-i\varepsilon, q_{0}^{2}) \Bigr]\!,
\end{equation}
where $\Delta p(q^2, q_{0}^{2}) = p(q^2) - p(q_{0}^{2})$.
By virtue of Eq.~(\ref{P_DQCD})
\begin{equation}
\label{Psc_DQCD}
\Delta p(s \pm i\varepsilon, q_0^2) =
\int_{m^2}^{\infty} \rho(\sigma)
\biggl[
\ln\biggl(\frac{s - \sigma \pm i\varepsilon}{s - m^2 \pm i\varepsilon}\biggr)\! +
\ln\biggl(\frac{m^2-q_0^2}{\sigma-q_0^2}\biggr)\!
\biggr]
\frac{d\,\sigma}{\sigma}.
\end{equation}
Then, since
\begin{equation}
\lim_{\varepsilon \to 0_{+}} \!\ln(x \pm i\varepsilon) = \ln |x| \pm i\pi\theta(-x),
\end{equation}
the first term in the square brackets of Eq.~(\ref{Psc_DQCD}) can
eventually be represented as (the limit $\varepsilon \to 0_{+}$ is
assumed hereinafter)
\begin{equation}
\ln\biggl(\!\frac{s - \sigma \pm i\varepsilon}{s - m^2 \pm i\varepsilon}\!\biggr)\!\! = \!
\ln\biggl| \frac{s - \sigma}{s - m^2} \biggr|\! \pm i\pi\theta(s-m^2)\theta(\sigma-s).
\end{equation}
Thus, Eq.~(\ref{Psc_DQCD}) acquires the form
\begin{equation}
\Delta p(s \pm i\varepsilon, q_0^2) =
\int_{m^2}^{\infty} \rho(\sigma)
\ln\biggl(
\biggl|\frac{s - \sigma}{s - m^2}\biggr|\frac{m^2-q_0^2}{\sigma-q_0^2}
\biggr)
\frac{d\,\sigma}{\sigma} \pm
i\pi\theta(s-m^2)\!\int_{s}^{\infty}\!\rho(\sigma)\frac{d\,\sigma}{\sigma},
\end{equation}
and, therefore, Eq.~(\ref{Rsc_DQCD}) reads
\begin{equation}
r(s) = \theta(s-m^2) \int_{s}^{\infty}\!
\rho(\sigma) \frac{d\,\sigma}{\sigma},
\end{equation}
that coincides with the integral representation~(\ref{R_DQCD}).

\end{document}